\documentclass[prd,a4paper,byrevtex]{revtex4}
\usepackage{graphicx}
\usepackage{dcolumn}
\usepackage{amsmath}
\usepackage{bm}

\begin{document}

\title{Equation of motion of an interstellar Bussard ramjet
with radiation and mass losses}

\author{Claude \surname{Semay}}
\thanks{FNRS Research Associate}
\email[E-mail: ]{claude.semay@umh.ac.be}
\affiliation{Groupe de Physique Nucl\'{e}aire Th\'{e}orique,
Universit\'{e} de Mons-Hainaut,
Acad\'{e}mie universitaire Wallonie-Bruxelles,
Place du Parc 20, BE-7000 Mons, Belgium}

\author{Bernard \surname{Silvestre-Brac}}
\email[E-mail: ]{silvestre@lpsc.in2p3.fr}
\affiliation{Laboratoire de Physique Subatomique et de Cosmologie,
Avenue des Martyrs 53, FR-38026 Grenoble-Cedex, France}

\date{\today}

\begin{abstract}
An interstellar Bussard ramjet is a spaceship using the protons of the
interstellar medium in a fusion engine to produce thrust. In recent
papers, it was shown that the relativistic equation of motion of an ideal
ramjet and of a ramjet with radiation loss are analytical. When a
mass loss appears, the limit speed of the ramjet is more strongly reduced. But, the
parametric equations, in terms of the ramjet's speed, for the position
of the ramjet in the inertial frame of the interstellar medium, the time
in this frame, and the proper time indicated by the clocks on board the
spaceship, can still be obtained in an analytical form. The
non-relativistic motion and the motion near the limit speed are studied.
\end{abstract}


\maketitle

\section{Introduction}
\label{sec:intro}

During its motion through the space, an interstellar fusion ramjet
collects interstellar ions (mostly protons) with a magnetic scoop (or
ramscoop) to supply a fusion reactor able to fuse protons to obtain
helium \cite{buss60,fish69,mall89,matl00}. Accelerated reaction products
are exhausted out of the spacecraft's rear to produce thrust. An
ideal ramjet could reach a
velocity very close to the speed of light $c$. If some energy extracted from the
interstellar medium is lost in form of thermal radiation, the ramjet speed 
is limited to a value below $c$ \cite{marx63}. 
In this two cases, we have shown that analytical formulas can be obtained
for the position of the ramjet in the inertial frame of the
interstellar medium, the time in this frame, and the proper time
indicated by the clocks on board the spaceship \cite{sema05,sema07}. These parametric
equations are given in terms of the ramjet's speed.

Moreover, it is natural to assume that a fraction of the collected interstellar 
gas can be lost during the work of the engine. The limit speed is then more 
strongly reduced \cite{marx63} and the equation of motion further more complicated.  
In this paper, we show that
analytical formulas can also be obtained for the position, the time and
the proper time of a ramjet with radiation and mass losses. The non-relativistic
motion and the motion near the limit speed are also studied.

The situation presented in this paper is more realistic than the case of a Bussard
ramjet with a perfect fusion engine \cite{sema05}. Nevertheless,
it is not very probable that the kind of ramjet considered here could be 
ever built, even in the far future. 
Consequently, this work can be considered as an advanced
exercise (from the point of view of calculus) in special relativity. The
framework of interstellar space travel could be very attractive for undergraduate students.
The basic equations are
simple outcomes of momentum and energy conservation. Even if the solutions demand
very heavy calculations, they have interesting properties to explore.
However, let us note that the ramjet concept is potentially too valuable to be simply
discarded despite its tremendous technical difficulties. Some
researchers have suggested alternatives to the initially proposed
proton-proton fusion ramjet \cite{mall89,matl00}: the catalytic ramjet
(the use of catalyzed fusion reaction with a high rate), the RAIR (the
use of nuclear fuel carried by the ship), etc. 

\section{General equations}
\label{sec:gen}

In the following, all calculations are performed in the frame of the
interstellar medium, considered as an inertial frame. A Bussard
ramjet of constant mass $M$ moves at speed $v=\beta c$ through this
medium, which contains protons at rest with a mass density $\rho$. The
effective intake area of the ramscoop is denoted $A$. A
fraction $\epsilon$ of the absorbed mass is converted into useful
kinetic energy in the hydrogen fusion reactor, a fraction $\lambda$
is dissipated in form of thermal radiation and a fraction $\kappa$ is lost in the 
interstellar medium (parameters $\epsilon$,
$\lambda$ and $\kappa$ have the same meaning than in Ref.~\cite{marx63}). This means
that, if a mass $dm$ of protons at rest is scooped up from the
interstellar medium, only an energy $\epsilon\, dm\, c^2$ is converted into
ordered motion of the exhausted material. 

In the frame of the spaceship,
we can assume that the thermal loss is isotropic and that the radiation
carries no momentum. For the ramjet, the energy absorbed due to the mass
$dm$ is $dE=dm\, \gamma\, c^2$, where $\gamma=1/\sqrt{1-\beta^2}$, and
the engine dissipates an energy equal to $\lambda\,dE$. Lorentz
transformations \cite{sear68,sema05b} imply that an energy
$\gamma\,\lambda\,dE=\lambda\, dm\, \gamma^2 c^2$ and a momentum
$\gamma\,\beta\,\lambda\,dE/c=\lambda\, dm\, \gamma^2 \beta\,c$ are lost
in the interstellar medium frame. 

If a fraction $\kappa$ of the collected interstellar gas is lost during the work of the engine, a
useless mass $\kappa\, dm$ is dropped out with no velocity in the frame of the ramjet. So, 
this mass has an energy $\kappa\, dm\, \gamma c^2$ and a momentum $\kappa\, dm\, \gamma \beta c$ in the 
interstellar medium. 

During the time interval $dt$, a mass $dm$ of protons, at rest, is
scooped-up. The ramjet speed is then increased by the quantity
$c d\beta$, thanks to the ejection of a mass $(1-\alpha)\,dm$
of helium with a speed $w c$, where 
\begin{equation}
\label{alpha}
\alpha=\epsilon+\lambda+\kappa.
\end{equation}
Another interesting quantity is the fraction of matter, 
\begin{equation}
\label{omega}
\omega=\epsilon+\lambda,
\end{equation}
which is converted into pure energy by the engine. 

The conservation of momentum implies that \cite{sear68,sema05b}
\begin{equation}
\label{convm}
M\,\gamma(\beta)\,\beta c = M\,\gamma(\beta+d\beta)\, (\beta+d\beta)c
+ \lambda\, dm\, \gamma(\beta)^2 \beta\,c + \kappa \, dm\, \gamma(\beta)
\beta\,c + (1-\alpha)\, dm\, \gamma(w)\, w c,
\end{equation}
where $\gamma(x)=1/\sqrt{1-x^2}$ and
$d\beta>0$. The conservation of energy leads to
\begin{equation}
\label{conve}
M\,\gamma(\beta)\,c^2 + dm\,c^2 = M\,\gamma(\beta+d\beta)\, c^2
+ \lambda\, dm\, \gamma(\beta)^2 c^2 + \kappa \, dm\, \gamma(\beta) c^2 +
(1-\alpha)\, dm\, \gamma(w)\, c^2.
\end{equation}
The collected mass is a function of the ramjet speed and is given by
\begin{equation}
\label{dm}
dm = A\, \rho\, \beta c \, dt.
\end{equation}

\section{Acceleration}
\label{sec:acc}

Taking into account equations~(\ref{convm})-(\ref{dm}), it is possible to 
compute the acceleration $\varphi=dv/dt$ of the ramjet, measured in the rest
frame of the interstellar medium. With this aim, it
is natural to define characteristic acceleration $\varphi_*$,
time $t_*$ and length $x_*$ by the
following relations
\begin{equation}
\label{characc}
\varphi_*=\frac{A\,\rho\,c^2}{M}, \quad t_*=\frac{c}{\varphi_*},
\quad x_*=\frac{c^2}{\varphi_*}=c t_*.
\end{equation}
It is also useful to introduce the following notations
\begin{equation}
\label{epsp}
u'=u\, (2-u), \quad \bar u = 1-u,
\end{equation}
with $1-u'=\bar u^2$.

It is worth noting that 
$\omega$ is only 0.0071 for the most energetic known
fusion reaction \cite{buss60}. So, for all fusion reactions, we have
$\omega'\approx 2\,\omega$ and $\bar \omega \approx 1$. If the particles
collected by the ramjet are an ideal mixing of matter and antimatter, the
mass reaction can be totally converted in pure energy ($\omega=1$ and
$\kappa=0$) \cite{sema05}.

Using the relation
\begin{equation}
\label{gamdbet}
\gamma(\beta+d\beta)=\gamma(\beta)+\gamma(\beta)^3\,\beta\,d\beta,
\end{equation}
the elimination of the reduced speed $w$ of the exhausted reaction mass by the
relation $\gamma(w)^2 (1-w^2)=1$ gives a second degree equation in $\varphi$,
whose physical solution is (in the following equations, the simplified
notation $\gamma$ means more precisely $\gamma(\beta)$)
\begin{align}
\label{phi}
\frac{\varphi}{\varphi_*}&=\frac{\sqrt{\gamma^2-1}}{\gamma^3}
\left( \sqrt{\gamma^2-1 + F(\gamma)} - \sqrt{\gamma^2-1}
\right) \quad \textrm{with} \\
F(\gamma)&=-\lambda'\, \gamma^2-2 \kappa\, \bar \lambda\, \gamma
+\kappa^2 +\alpha' .
\end{align}
This equation gives the acceleration
of the ramjet as a function of its speed in the inertial frame of the
interstellar medium. If $\alpha=0$ ($\epsilon=\lambda=\kappa=0$), the
acceleration vanishes as there is no input of energy into the engine.
Since $\varphi=0$ if $\beta=0$ ($\gamma=1$), an initial boosting is necessary for the
ramjet. This is due to the fact that the reaction mass reaches the
reactor thanks to the speed of the ramjet. In theory, a very small speed
is sufficient to start the ramjet. In practice, a fusion reactor could
probably not operate correctly without a sufficient intake.
Nevertheless,
the ramjet could accelerate with an initial speed as low as 10~km/s
\cite{buss60}. Such a speed could be reached with usual chemical rockets
or by future nuclear rockets. As expected, the acceleration also vanishes for $\beta=1$
($\gamma=\infty$) since no object can move at the speed of light.

\begin{center}
\begin{figure}
\includegraphics*[height=7cm]{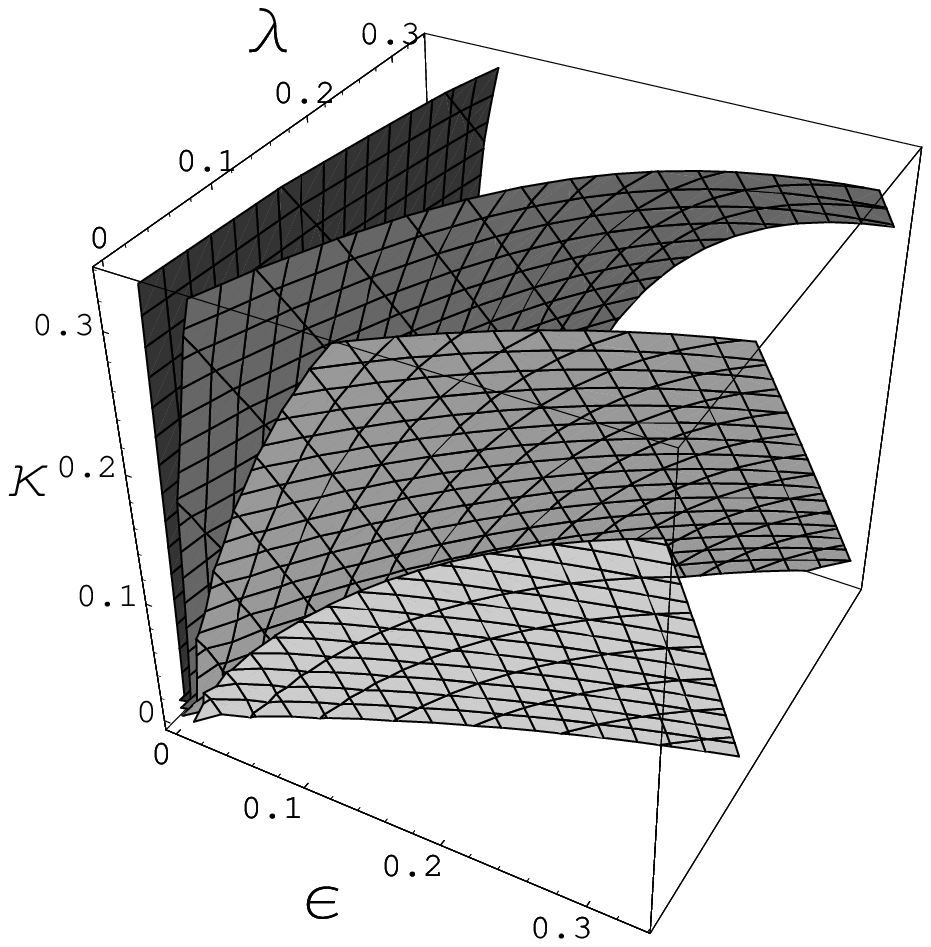}
\includegraphics*[height=7cm]{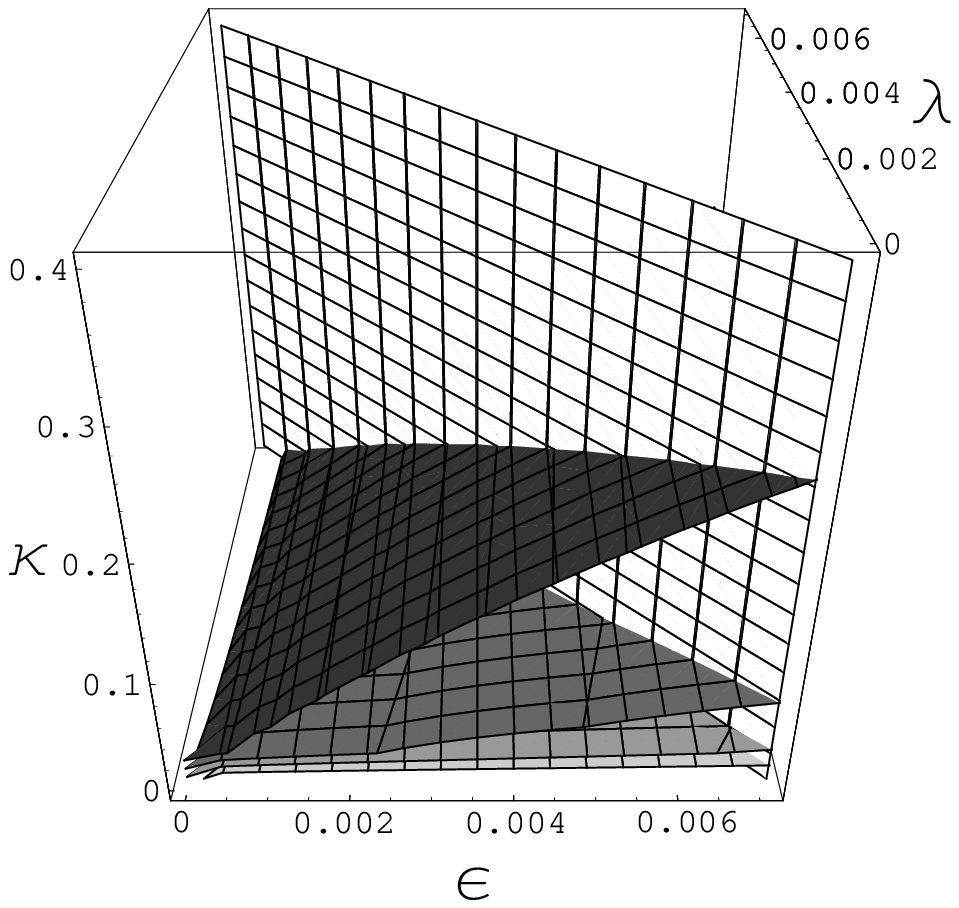}
\caption{Limit reduced speed $\beta_l$ of the ramjet as a function of the parameters
$\epsilon$, $\lambda$ and $\kappa$. Surfaces colored in dark grey to light grey correspond
respectively to $\beta_l=0.2$, 0.4, 0.6, 0.8. Left: Arbitrary values of $\epsilon$, 
$\lambda$ and $\kappa$. Right: Values of $\omega \leq 0.0071$ with 
the vertical white plane marking the constraint $\omega = 0.0071$.
\label{fig:betal}}
\end{figure}
\end{center}

But this limit speed is never reached for a non-ideal ramjet. It is clear from Eq.~(\ref{phi})
that $\varphi=0$ when $F(\gamma)=0$. $F(\gamma)$ is a quadratic function in $\gamma$ with two roots
$-\Gamma_l$ and $\gamma_l$ such that
\begin{align}
\label{fgamma}
F(\gamma)&=\lambda'(\Gamma_l+\gamma)(\gamma_l-\gamma) \quad \textrm{with} \\
\label{gammal}
\gamma_l&=\frac{\sqrt{\kappa^2+\lambda' \alpha'}-\bar \lambda \kappa}{\lambda'} \quad \textrm{and} \quad
\Gamma_l=\frac{\sqrt{\kappa^2+\lambda' \alpha'}+\bar \lambda \kappa}{\lambda'}.
\end{align}
With a little algebra, it can be shown that
\begin{equation}
\label{gammal2}
1 < \gamma_l \leq \Gamma_l,
\end{equation}
taking into account the condition $\alpha < 1$ to obtain a non-vanishing exhausted mass.
For relevant values of the speed ($\gamma \geq 1$), one can see that: $F(\gamma_l) =0$,
$F(\gamma) \geq 0$ for $\gamma \in [1,\gamma_l]$ and 
$F(\gamma) \leq 0$ for $\gamma \geq \gamma_l$. So, $\gamma_l$ is a limit value: 
if $\gamma = \gamma_l$, the ramjet acceleration vanishes ($\varphi = 0$);
if $\gamma < \gamma_l$, the ramjet speed increases ($\varphi > 0$); 
if $\gamma > \gamma_l$, the ramjet speed decreases ($\varphi < 0$).
Thus, the ramjet speed tends toward
the limit reduced speed $\beta_l= \sqrt{1-\frac{1}{\gamma_l^2}}$. In the following, 
we will only consider the realistic case $\beta < \beta_l$ or $\gamma < \gamma_l$.
Contrary to the case of an
ideal ramjet, the velocity of a ramjet with radiation and/or mass losses cannot be
arbitrarily close to the speed of light \cite{sema05} (see Figs.~\ref{fig:betal} and
\ref{fig:betat}).  

Even with small values of the parameters $\lambda$ and $\kappa$,
the maximum speed can differ significantly from 1.
This means that the slowing down of time on board the spaceship can
become not large enough to allow interstellar travels in a period
of time bearable for human beings. For values of $\lambda+\kappa \gg \epsilon$,
the motion of the ramjet becomes non-relativistic.

In our previous papers \cite{sema05,sema07}, all equations were given as function
of the reduced speed $\beta$. We find here more convenient to present the
parametric equations of motion of the ramjet as a function of its reduced speed 
through the kinematical factor $\gamma = \gamma(\beta)$. We will assume
that, at a time $t=0$ in the inertial frame, the position of the ramjet
is $x=0$ and its reduced speed is $\beta_0 \ne 0$. Moreover, the clocks
on board the ramjet indicate a proper time $\tau=0$.

\section{Time}
\label{sec:time}

\begin{center}
\begin{figure}
\includegraphics*[height=7cm]{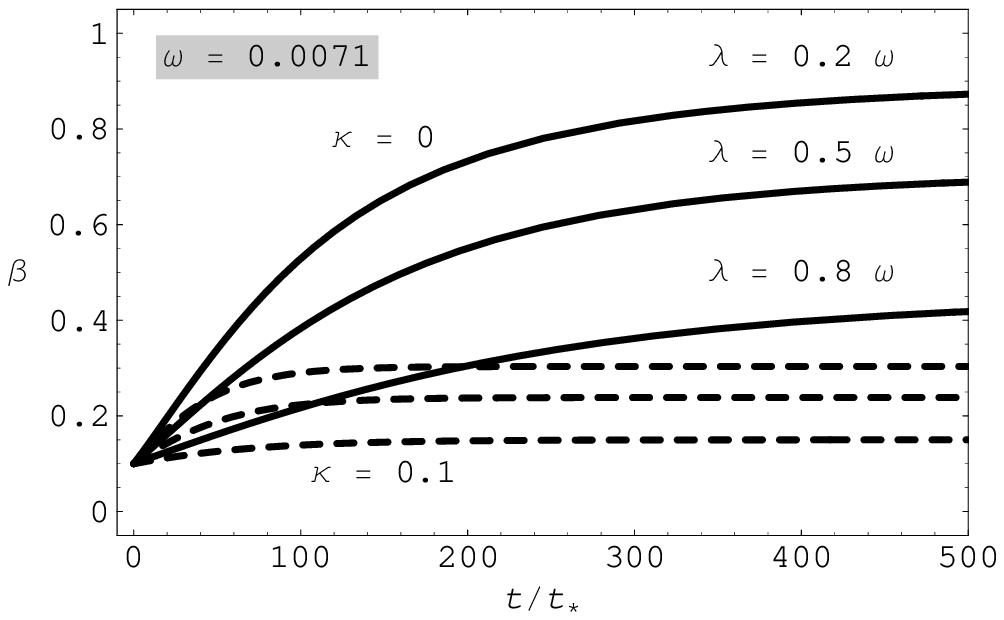}
\caption{Reduced speed $\beta$ of the ramjet as a function of the
reduced time $t/t_*$ spent in the inertial frame of the
interstellar medium, for the parameter $\omega=0.0071$ \cite{buss60} and an
initial reduced speed $\beta_0=0.1$. The solid (dashed) curve corresponds to $\kappa=0$ ($\kappa=0.1$).
For each value of $\kappa$, curves from bottom to top are respectively drawn for values 0.8, 0.5, 0.2 of the ratio
$\lambda/\omega$.}
\label{fig:betat}
\end{figure}
\end{center}

Since 
\begin{equation}
\label{tphi}
\frac{d\gamma}{dt} = \gamma^3\,\beta\frac{d\beta}{dt}
= \gamma^2 \sqrt{\gamma^2-1} \frac{\varphi}{c},
\end{equation}
the solution of Eq.~(\ref{phi}) is given by the following integral
\begin{equation}
\label{intdbt}
\frac{1}{t_*}\int_0^t dt = \int_{\gamma_0}^\gamma \frac{\gamma \, d\gamma}
{(\gamma^2-1)\left( \sqrt{\gamma^2-1 + F(\gamma)} - \sqrt{\gamma^2-1}
\right)}.
\end{equation}
We give here the main steps of the procedure to solve this integral:
\begin{itemize}
  \item To multiply the numerator and the denominator of the fraction by the quantity
  ($\sqrt{\gamma^2-1 + F(\gamma)} + \sqrt{\gamma^2-1}$) to obtain 
  the function $(\gamma^2-1)F(\gamma)$ at the denominator;
  \item Using the relation~(\ref{fgamma}), to write the fraction obtained as a sum of simpler fractions;
  \item To integrate each of these new fractions and to simplify the result.
\end{itemize}
The calculation is very heavy, but an analytical form can be found. To get a concise writing, 
it is useful to define some intermediate quantities:
\begin{align}
\label{interm1}
\Lambda_1(z)&=\bar \lambda - z, \\
\Lambda_2(z)&=\bar \lambda z - \kappa, \\
U(z,\gamma)&=\sqrt{(z^2-1)(\gamma^2-1)}+z \gamma-1, \\
V(z,\gamma)&=2\sqrt{(\Lambda_2^2(z)-\bar \alpha^2)(\Lambda_2^2(\gamma)-\bar \alpha^2)}
+2\bar \lambda \Lambda_2(z) \gamma -2 (\kappa \Lambda_2(z) + \bar \alpha^2), \\
W_\pm(z,\gamma)&=2\sqrt{(\Lambda_1^2(z)-\bar \alpha^2)(\Lambda_2^2(\gamma)-\bar \alpha^2)}
\pm 2\bar \lambda \Lambda_1(z) \gamma -2 (z \Lambda_1(z) + \bar \alpha^2), \\
S_\pm(z,\gamma,\gamma_0)&=\sqrt{\Lambda_1^2(z)-\bar \alpha^2}
\ln \frac{(\gamma_0\mp 1) W_\pm(z,\gamma)}{(\gamma\mp 1) W_\pm(z,\gamma_0)} , \\
R(z,\gamma,\gamma_0)&=\frac{1}{\sqrt{z^2-1}} 
\ln \frac{(z-\gamma_0) U(z,\gamma)}{(z-\gamma) U(z,\gamma_0)} 
+ \frac{\sqrt{\Lambda_2^2(z)-\bar \alpha^2}}{z^2-1} 
\ln \frac{(z-\gamma_0) V(z,\gamma)}{(z-\gamma) V(z,\gamma_0)} .
\label{interm2}
\end{align}
We can then write
\begin{equation}
\label{t}
\frac{t}{t_*}=\frac{\gamma_l R(\gamma_l,\gamma,\gamma_0) + \Gamma_l R(-\Gamma_l,\gamma,\gamma_0)}
{\lambda'(\gamma_l+\Gamma_l)}
-\frac{S_+(\kappa,\gamma,\gamma_0)}{2\lambda'(\gamma_l-1)(\Gamma_l+1)}
-\frac{S_-(-\kappa,\gamma,\gamma_0)}{2\lambda'(\gamma_l+1)(\Gamma_l-1)}.
\end{equation}
We can see on Fig.~\ref{fig:betat}, how the speed of the ramjet tends toward the limit speed as time increases.

It is interesting to look at the limit $\kappa\rightarrow 0$. In this case, 
we have:
\begin{align}
\label{tlimit1}
\gamma_l &\stackrel{\kappa = 0}{\longrightarrow} \sqrt{\frac{\alpha'}{\lambda'}}, \\
\beta_l &\stackrel{\kappa = 0}{\longrightarrow} \sqrt{\frac{\alpha'-\lambda'}{\alpha'}}, \\
\sqrt{\Lambda_1^2(\kappa)-\bar \alpha^2} &\stackrel{\kappa = 0}{\longrightarrow} \sqrt{\alpha'-\lambda'}, \\
\sqrt{\Lambda_2^2(\gamma_l)-\bar \alpha^2} &\stackrel{\kappa = 0}{\longrightarrow} \sqrt{\frac{\alpha'-\lambda'}{\lambda'}},
\label{tlimit2}
\end{align}
where $\alpha = \epsilon+\lambda$ in the r.h.s.\ of these relations.
Moreover, $\Gamma_l$ and $\gamma_l$ tend toward the same limit. In formula~(\ref{t}), one can see that, in the limit $\kappa\rightarrow 0$, the coefficient of the functions $R$ tends toward $\frac{\gamma_l}{2 \alpha' \beta_l}$ and the coefficient of the functions $S_\pm$ tends toward $\frac{1}{2 \sqrt{\alpha'} \beta_l}$, where parameters $\gamma_l$ and $\beta_l$ are taken for $\kappa=0$.
So, one can see that expressions for the function $t/t_*$ obtained in this paper when the parameter $\kappa$ vanishes
and given by formula~(19) in Ref.~\cite{sema07} are identical
(the supplementary factor $1/2$ can be absorbed in the $\ln$-functions).

\section{Proper time}
\label{sec:tau}

\begin{center}
\begin{figure}
\includegraphics*[height=7cm]{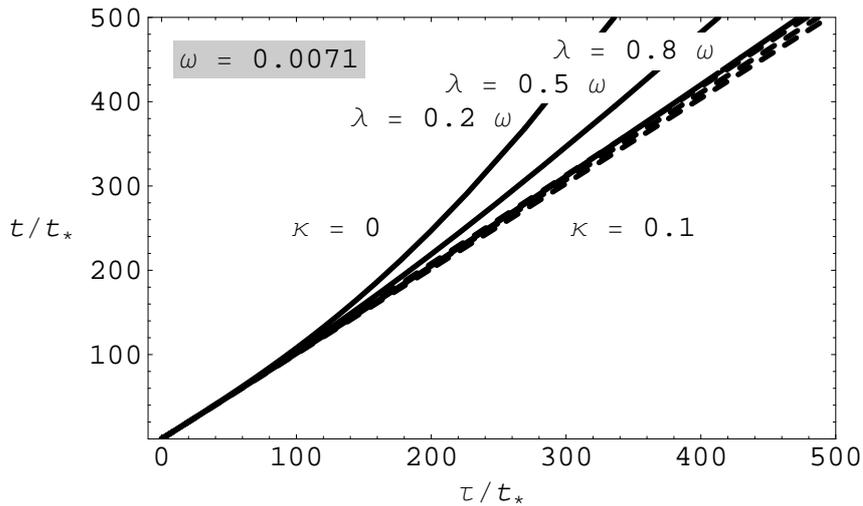}
\caption{Reduced time $t/t_*$ spent in the inertial frame of the
interstellar medium as a function of the reduced proper time $\tau/t_*$
on board the ramjet, for the parameter $\omega=0.0071$ \cite{buss60} and an
initial reduced speed $\beta_0=0.1$. The solid (dashed) curve corresponds to $\kappa=0$ ($\kappa=0.1$).
For each value of $\kappa$, curves from bottom to top are respectively drawn for values 0.8, 0.5, 0.2 of the ratio
$\lambda/\omega$.}
\label{fig:ttau}
\end{figure}
\end{center}

Using the well known relation between the time and the proper time
$d\tau=dt/\gamma$ \cite{sear68,sema05b}, Eq.~(\ref{intdbt}) simplifies and
the proper time is given by the following integral
\begin{equation}
\label{intdbtau}
\frac{1}{t_*}\int_0^\tau d\tau = \int_{\gamma_0}^\gamma \frac{d\gamma}
{(\gamma^2-1)\left( \sqrt{\gamma^2-1 + F(\gamma)} - \sqrt{\gamma^2-1}
\right)}.
\end{equation}
An analytical solution of this integral can be found with a procedure similar 
to the one used for the calculation of $t/t_*$. Using again the 
notations~(\ref{interm1})-(\ref{interm2}), a tedious calculation gives
\begin{equation}
\label{tau}
\frac{\tau}{t_*}=\frac{R(\gamma_l,\gamma,\gamma_0) - R(-\Gamma_l,\gamma,\gamma_0)}
{\lambda'(\gamma_l+\Gamma_l)}
-\frac{S_+(\kappa,\gamma,\gamma_0)}{2\lambda'(\gamma_l-1)(\Gamma_l+1)}
+\frac{S_-(-\kappa,\gamma,\gamma_0)}{2\lambda'(\gamma_l+1)(\Gamma_l-1)}.
\end{equation}
The link between the proper time on board the spaceship and the
time spent in the inertial frame of the interstellar medium can then be
computed (see Fig.~\ref{fig:ttau}).

Using relations~(\ref{tlimit1})-(\ref{tlimit2}) in formula~(\ref{tau}), one can see that, in the limit $\kappa\rightarrow 0$, the coefficient of the functions $R$ tends toward $\frac{1}{2 \alpha' \beta_l}$ and the coefficient of the functions $S_\pm$ tends toward $\frac{1}{2 \sqrt{\alpha'} \beta_l}$, where parameters $\gamma_l$ and $\beta_l$ are taken for $\kappa=0$.
Again, one can see that expressions for the function $\tau/t_*$ obtained in this
paper when the parameter $\kappa$ vanishes and given by formula~(24) in Ref.~\cite{sema07}
are identical (the supplementary factor $1/2$ can be absorbed in the $\ln$-functions).

\section{distance}
\label{sec:dist}

\begin{center}
\begin{figure}
\includegraphics*[height=7cm]{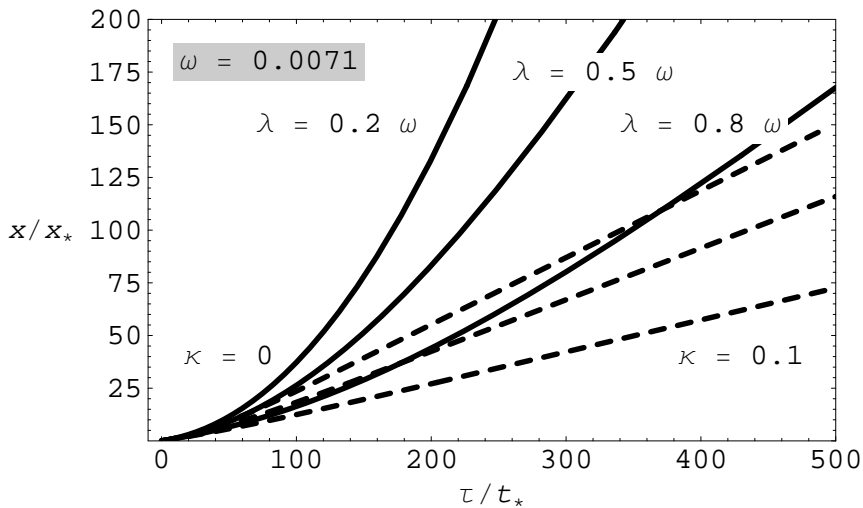}
\caption{Reduced distance $x/x_*$ travelled by the ramjet in the
inertial frame of the interstellar medium as a function of the reduced
proper time $\tau/t_*$ on board the ramjet, for the parameter $\omega=0.0071$ \cite{buss60} and an
initial reduced speed $\beta_0=0.1$. The solid (dashed) curve corresponds to $\kappa=0$ ($\kappa=0.1$).
For each value of $\kappa$, curves from bottom to top are respectively drawn for values 
0.8, 0.5, 0.2 of the ratio $\lambda/\omega$.}
\label{fig:xtau}
\end{figure}
\end{center}

Using the relations
\begin{equation}
\label{xphi}
c dt=\frac{dx}{\beta}=\frac{\gamma\, dx}{\sqrt{\gamma^2-1}},
\end{equation}
Eq.~(\ref{intdbt}) can be rewritten into the form
\begin{equation}
\label{intdbx}
\frac{1}{x_*}\int_0^x dx = \int_{\gamma_0}^\gamma \frac{d\gamma}
{\sqrt{\gamma^2-1}\left( \sqrt{\gamma^2-1 + F(\gamma)} - \sqrt{\gamma^2-1}
\right)}.
\end{equation}
Again, an analytical solution of this integral can be found with a procedure similar 
to the one used for the calculation of $t/t_*$ and $\tau/t_*$. 
New intermediate quantities must be defined:
\begin{align}
\label{interm3}
L&=\bar \lambda - \frac{\epsilon}{2}, \\
A&=\bar \alpha + \frac{\epsilon}{2}, \\
Q&=\frac{\bar \lambda\, \bar \alpha}{L \, A} \\
I_\pm(z)&=\frac{\bar \lambda z \pm (\bar \alpha-\kappa)}{z\mp 1}, \\
J_\pm(z)&=\frac{\bar \lambda z \pm (\bar \alpha+\kappa)}{z\pm 1}, \\
\nu(z)&=\arcsin \sqrt{\frac{L}{J_-(z)}}. 
\label{interm4}
\end{align}
We can then write
\begin{align}
\label{x}
\frac{x}{x_*}&=X(\gamma)-X(\gamma_0) \quad \textrm{with} \quad \\
X(\gamma)&=\frac{1}{\lambda'(\gamma_l+\Gamma_l)} \left[
\ln \frac{\Gamma_l+\gamma}{\gamma_l-\gamma}+ 
\frac{\epsilon\, I_+(\gamma_l)}{\sqrt{L \, A}} 
\Pi \left( \frac{J_-(\gamma_l)}{L}, \nu(\gamma), Q \right)
-\frac{\epsilon\, I_-(\Gamma_l)}{\sqrt{L \, A}} 
\Pi \left( \frac{J_+(\Gamma_l)}{L}, \nu(\gamma), Q \right)
\right] ,
\label{x2}
\end{align}
where $\Pi$ is the incomplete elliptic integral \cite{grad80}.
The distance travelled by the ramjet in the interstellar medium as a
function of the proper time indicated by the on board clocks can then
be computed (see Fig.~\ref{fig:xtau}).

In the limit $\kappa\rightarrow 0$, the first term of formula~(\ref{x2}) reduces to
$\frac{\gamma_l}{\alpha'}\arg\tanh \frac{\gamma}{\gamma_l}$, 
where parameters $\gamma_l$ and $\beta_l$ are taken for $\kappa=0$.
Again, one can see that expressions for the function $x/x_*$ obtained in this paper
when the parameter $\kappa$ vanishes and given by formula~(32) in Ref.~\cite{sema07}
are identical.

\section{Non-relativistic limit}
\label{sec:nr}

In the non-relativistic limit ($\beta \ll 1$), Eq.~(\ref{phi}) reduces to
\begin{equation}
\label{phinr}
\frac{\varphi}{\varphi_*} \approx \sqrt{F(1)} \beta,
\end{equation}
where
\begin{equation}
\label{f1}
F(1)=\epsilon (2 \bar \alpha + \epsilon).
\end{equation}
This number is clearly positive since $\epsilon > 0$ and $\alpha < 1$. Eq.~(\ref{phinr})
can be integrated to give
\begin{align}
\label{phinr2}
\frac{x}{x_*}&=\frac{\beta-\beta_0}{\sqrt{\epsilon (2 \bar \alpha + \epsilon)}}\quad \textrm{with} \\
\beta&=\beta_0 \exp \left( \sqrt{\epsilon (2 \bar \alpha + \epsilon)} \frac{t}{t_*} \right).
\end{align}
These expressions, characteristic of an exponential motion \cite{sema05,sema07}, are valid
provided $t \ll -t_* \ln \beta_0/\sqrt{\epsilon (2 \bar \alpha + \epsilon)}$. 
Let us remember that $\tau$ does not differ 
significantly from $t$ at non-relativistic speed.

\section{Asymptotic motion}
\label{sec:asymp}

Near the limit speed, the term $F(\gamma)$ is very small with respect to the term $\gamma^2-1$, 
so we can write
\begin{equation}
\label{Fsmall}
\sqrt{\gamma^2-1 + F(\gamma)} - \sqrt{\gamma^2-1} \approx
\frac{F(\gamma)}{2 \sqrt{\gamma^2-1}}.
\end{equation}
Equation~(\ref{phi}) reduces then to
\begin{equation}
\label{phiasymp}
\frac{\varphi}{\varphi_*} \approx \frac{F(\gamma)}{2\gamma^3}.
\end{equation}
It is worth noting that, in the case of an ideal ramjet ($\lambda=\kappa=0$), $F(\gamma)=\epsilon'$ and
Eq.~(\ref{phiasymp}) is then the equation of motion of an uniformly accelerated spaceship with a constant proper
acceleration $\varphi_* \epsilon'/2$ \cite{sema05}.

If the ramjet is characterized by a mass $M=1000$~t and an effective intake area $A=10^4$~km$^2$, 
the characteristic acceleration is $\varphi_*\approx 1500$~m/s$^2$ for an optimistic value 
 of the particle density, let us say $10^3$/cm$^3$ \cite{buss60}.
Within such conditions, the asymptotic proper acceleration of an ideal ramjet with 
$\epsilon=0.0071$ is about 1~g. With an initial speed as low as 10~km/s, 
the spacecraft could then approach light velocity within a year.
With a more realistic value, 1000 times smaller, for the particle density, 
the performance of the ramjet would be reduced by the same factor. 

In the asymptotic regime, the reduced speed is very close the reduced limit speed.
So we consider values of $\beta$ such that $\beta_s \leq \beta < \beta_l$, with
$\beta_l - \beta_s \ll \beta_l$. Using Eq.~(\ref{gamdbet}), 
the factor $\gamma_s=\gamma(\beta_s)$ can then be approximately written
\begin{equation}
\label{gamasymp}
\gamma_s = \gamma(\beta_l-(\beta_l-\beta_s))\approx \gamma_l-\gamma_l^3 \beta_l (\beta_l-\beta_s).
\end{equation}
So, we have
\begin{equation}
\label{betaasymp}
\beta_l -\beta_s \approx \frac{1}{\beta_l \gamma_l^3} (\gamma_l - \gamma_s).
\end{equation}
The condition $\beta_l - \beta_s \ll \beta_l$ is then equivalent to
$\gamma_l - \gamma_s \ll \gamma_l^3 \beta_l^2 < \gamma_l^3$. 
This inequality is relevant since $\gamma_l$ has a finite value when $\lambda\not = 0$ or $\kappa\not = 0$.
We assume that the ramjet is at position $x_s$, at time $t_s$, and
at proper time $\tau_s$
when it has a speed $\beta_s$. Using the approximation~(\ref{Fsmall}), the integral for
the position is given by
\begin{equation}
\label{xs1}
\frac{1}{x_*}\int_{x_s}^x dx \approx \int_{\gamma_s}^\gamma \frac{2 d\gamma}
{\lambda'(\Gamma_l+\gamma)(\gamma_l-\gamma)} \approx \frac{2}{\lambda'(\gamma_l+\Gamma_l)}
\int_{\gamma_s}^\gamma \frac{d \gamma}{\gamma_l-\gamma},
\end{equation}
since $\gamma \approx \gamma_l$ under the integral. The corresponding solution is
\begin{equation}
\label{gammaxs}
\gamma = \gamma_l -(\gamma_l - \gamma_s)\exp\left( -\frac{\lambda' (\gamma_l + \Gamma_l)}
{2x_*}(x-x_s)\right).
\end{equation}
One can treat the time exactly in the same way. Remarking that $\gamma/\sqrt{\gamma^2-1}
\approx 1/\beta_l$ under the integral, the solution is given by
\begin{equation}
\label{gammats}
\gamma = \gamma_l -(\gamma_l - \gamma_s)\exp\left( -\frac{\lambda' \beta_l(\gamma_l + \Gamma_l)}
{2t_*}(t-t_s)\right).
\end{equation}
This equation can be directly deduced from equation~(\ref{gammaxs}), since
$x-x_s \approx \beta_l c (t-t_s)$ in this regime.
Similar calculations for the proper time lead to
\begin{equation}
\label{gammataus}
\gamma = \gamma_l -(\gamma_l - \gamma_s)\exp\left( -\frac{\lambda' \beta_l \gamma_l
(\gamma_l + \Gamma_l)}{2t_*}(\tau-\tau_s)\right).
\end{equation}
This equation can be directly deduced from equation~(\ref{gammats}), since
$t-t_s \approx \gamma_l (\tau-\tau_s)$ in this regime.

Since $\beta_s \leq \beta <\beta_l$, Eq.~(\ref{betaasymp}) holds also for $\beta_s$ replaced
by $\beta$.
Consequently, in Eqs.~(\ref{gammaxs})-(\ref{gammataus}), quantities $\gamma$, $\gamma_l$, $\gamma_s$ 
outside the exponentials can be replaced respectively by $\beta$, $\beta_l$, $\beta_s$.
In the limit $\kappa \rightarrow 0$, it can be checked that Eqs.~(\ref{gammaxs}), (\ref{gammats}),
(\ref{gammataus}) tend respectively towards formulas~(31), (17), (23) in Ref.~\cite{sema07}
(note a misprint in Eq.~(31): the factor $\beta_l$ must be suppressed in the exponential).

\section{Fundamental inequalities}
\label{sec:ineq}

Since $\gamma > 1$, we have
\begin{equation}
\label{ineqt}
\tau = \int_0^\tau d\tau =\int_0^t \frac{dt}{\gamma} < \int_0^t dt = t.
\end{equation}
Moreover, since $\beta < 1$, we have
\begin{equation}
\label{ineqx}
x = \int_0^x dx =\int_0^t \beta c\, dt < \int_0^t c\, dt = c\,t.
\end{equation}
So, from these inequalities, we can conclude that functions~(\ref{t}), (\ref{tau}) 
and (\ref{x}) are characterized by
\begin{align}
\label{ineqtx}
\frac{\tau}{t_*}(\gamma,\gamma_0) &< \frac{t}{t_*}(\gamma,\gamma_0), \\
\frac{x}{x_*}(\gamma,\gamma_0) &< \frac{t}{t_*}(\gamma,\gamma_0).
\end{align}
It is not evident to demonstrate these properties from the explicit form
of these functions, except in the case of a perfect antimatter ramjet
($\epsilon=1$, $\lambda=\kappa=0$) \cite{sema05}. Nevertheless, we have checked
that they are fulfilled numerically in any case.

\section{Summary}
\label{sec:summary}

Formulas~(\ref{t}), (\ref{tau}) and (\ref{x}) form the complete set of
parametric equations of motion for a Bussard ramjet with radiation and mass losses,
as a function of its speed. It is then easy to compute, for instance,
the distance traveled by the ramjet in the interstellar medium as a
function of the proper time indicated by the on board clocks
(see Fig.~\ref{fig:xtau}), or the link between this proper time and the
time spent in the inertial frame of the interstellar medium
(see Fig.~\ref{fig:ttau}).

With radiation and mass losses, the ramjet speed cannot be arbitrarily close to the
speed of light. A limit speed, only reached asymptotically (see
Figs.~\ref{fig:betal} and \ref{fig:betat}), lowers the
performance of a Bussard ramjet as an interstellar spaceship or as a
time machine for the exploration of the future \cite{nahi99}. 

\section{Acknowledgments}

C. Semay would like to thank the FNRS for financial support.


\end{document}